\documentclass[english,aps,prl,reprint,superscriptaddress]{revtex4-1}
\usepackage[T1]{fontenc}
\usepackage[latin9]{inputenc}
\usepackage{babel}
\usepackage{amsmath}
\usepackage{amssymb}
\usepackage{graphicx}
\usepackage[unicode=true,pdfusetitle,
 bookmarks=true,bookmarksnumbered=false,bookmarksopen=false,
 breaklinks=false,pdfborder={0 0 0},backref=false,colorlinks=false]
 {hyperref}

\makeatletter
\@ifundefined{textcolor}{}
{%
 \definecolor{BLACK}{gray}{0}
 \definecolor{WHITE}{gray}{1}
 \definecolor{RED}{rgb}{1,0,0}
 \definecolor{GREEN}{rgb}{0,1,0}
 \definecolor{BLUE}{rgb}{0,0,1}
 \definecolor{CYAN}{cmyk}{1,0,0,0}
 \definecolor{MAGENTA}{cmyk}{0,1,0,0}
 \definecolor{YELLOW}{cmyk}{0,0,1,0}
 }

\usepackage{babel}
\usepackage{times}
\usepackage{braket}
\usepackage{txfonts}

\makeatother

\begin{document}

\title{Quantum discord cannot be shared}

\author{Alexander Streltsov}

\email{streltsov@thphy.uni-duesseldorf.de}

\affiliation{Heinrich-Heine-Universität Düsseldorf, Institut für Theoretische
Physik III, D-40225 Düsseldorf, Germany}

\affiliation{Theoretical Division, MS-B213, Los Alamos National Laboratory, Los
Alamos, New Mexico 87545, USA}

\author{Wojciech H. Zurek }

\affiliation{Theoretical Division, MS-B213, Los Alamos National Laboratory, Los
Alamos, New Mexico 87545, USA}

\affiliation{Santa Fe Institute, 1399 Hyde Park Rd., Santa Fe, New Mexico 87501,
USA}
\begin{abstract}
Niels Bohr proposed that the outcome of the measurement becomes objective
and real, and, hence, classical, when its results can be communicated
by classical means. In this work we revisit Bohr's postulate using
modern tools from the quantum information theory. We find a full confirmation
of Bohr's idea: if a measurement device is in a nonclassical state,
the measurement results cannot be communicated perfectly by classical
means. In this case some part of information in the measurement apparatus
is lost in the process of communication: the amount of this lost information
turns out to be the quantum discord. The information loss occurs even
when the apparatus is \emph{not} entangled with the system of interest.
The tools presented in this work allow to generalize Bohr's postulate:
we show that for pure system-apparatus states quantum communication
does not provide any advantage when measurement results are communicated
to more than one recipient. We further demonstrate the superiority
of quantum communication to two recipients on a mixed system-apparatus
state and show that this effect is fundamentally different from quantum
state cloning.
\end{abstract}
\maketitle
Quantum measurement problem arises because our Universe is quantum
to the core, so the interaction between the (quantum) apparatus ${\cal A}$
and the measured quantum system ${\cal S}$ creates correlation that
is also quantum. The joint states of the apparatus and the system
can be even entangled (as is illustrated by the Schrödinger's cat
\cite{Schroedinger1935}) which leads to interpretational problems
such as the basis ambiguity \cite{Zurek1981}.

Quantum entanglement is a hallmark of quantum correlations: Pure states
that possess it violate Bell's inequality, showing that quantum mechanics
contradicts classical intuition resting on the assumption (expressed
by Einstein, Podolsky, and Rosen \cite{Einstein1935}) that state
is local -- a property of an individual system. Yet, as our daily
experience convincingly demonstrates, Einstein's intuition about reality
is respected on the macroscopic level: A state is a property of an
individual system, and while correlations exist, a completely known
state of composite classical system can be always expressed as a ``Cartesian
product'' of pure local states, so that a state of each component
is also completely known. Entanglement depends on tensor structure
of composite quantum states that allows and even mandates \cite{Zurek2003a,Zurek2005,Zurek2011}
ignorance of the ingredients for a completely known composite state.

Entanglement is simply impossible in the classical realm. In general
(for mixed -- i.e., incompletely known -- states) quantum entanglement
is defined by specifying how a composite state can be put together.
When it can be assembled from classical ingredients by mixing direct
(ultimately Cartesian) products of quantum states of individual subsystems,
composite state can be prepared by observers that employ only local
operations and classical communication (LOCC). Such states are known
as separable, and, by definition, they are not entangled \cite{Werner1989}. 

Here we show that quantum discord can be defined by considering the
opposite of the process of assembling a state. That is, when one attempts
to pull apart a quantum state so that, in the end, all the ingredients
are classical and can be communicated classically to distant recipients,
the cost of such an operation is given by quantum discord. Thus, discord
is the information lost when a composite quantum state is disassembled.
It is amusing to note that this disparity between how hard it is to
pull a state apart compared to how difficult it is to put it together
has the opposite ``sign'' than in everyday experience, as it is
generally easier to take apart a device -- e.g. a clock -- than it
is to put it back together.

Quantum discord is the difference between the quantum mutual information
$I({\cal S}:{\cal A})=S(\rho^{{\cal S}})+S(\rho^{{\cal A}})-S(\rho^{{\cal SA}})$
and the information $J$ accessible via the measurement with outcomes
$\{E_{i}^{{\cal A}}\}$, record states of the apparatus ${\cal A}$:
$J({\cal S}:{\cal A})_{\left\{ E_{i}^{{\cal A}}\right\} }=S(\rho^{{\cal S}})-\sum_{i}p_{i}S(\rho_{i}^{{\cal S}})$,
where $S$ is the von Neumann entropy, $p_{i}=\mathrm{Tr}[E_{i}^{{\cal A}}\rho^{{\cal SA}}]$
is the probability of outcome $i$, and $\rho_{i}^{{\cal S}}$ is
the state of the system after the outcome $i$ has been obtained:
$\rho_{i}^{{\cal S}}=\mathrm{Tr}_{{\cal A}}[E_{i}^{{\cal A}}\rho^{{\cal SA}}]/p_{i}$.
In the classical domain $I$ and $J$ coincide as there is an underlying
joint probability distribution that can be used to express the joint
state of the two systems in terms of the local states of individual
subsystems. In that case, Bayes' rule holds, and $I$ is identically
equal to $J$ \cite{Cover1991}. However, in quantum physics obtaining
conditional information requires a measurement, and that generally
alters the measured state, so the information obtained by local measurements
is less than the mutual information present in the joint pre-measurement
state. The difference between the mutual information $I$ and the
information accessible via the measurement apparatus $J$ is known
as quantum discord $\delta({\cal S}:{\cal A})_{\{E_{i}^{{\cal A}}\}}=I({\cal S}:{\cal A})-J({\cal S}:{\cal A})_{\{E_{i}^{{\cal A}}\}}$
\cite{Zurek2000,Ollivier2001,Henderson2001,Modi2012}. Quantum discord
is then at least as large as its minimum $\delta({\cal S}:{\cal A})=\min_{\{E_{i}^{{\cal A}}\}}\delta({\cal S}:{\cal A})_{\{E_{i}^{{\cal A}}\}}$.

\begin{figure}
\begin{centering}
\includegraphics[width=1\columnwidth]{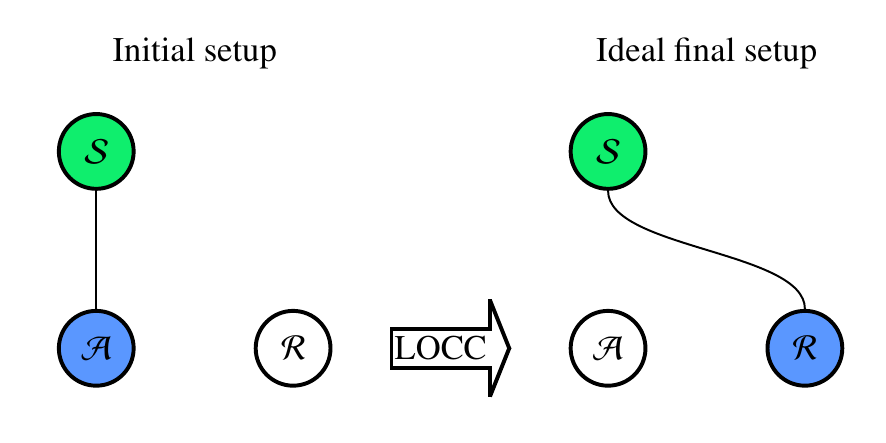} 
\par\end{centering}

\caption{\label{fig:LOCC}Ideal classical communication to a distant recipient.
Left part of the figure shows the initial situation: the system ${\cal S}$
is correlated with the apparatus ${\cal A}$. The recipient ${\cal R}$
is initially not correlated with ${\cal SA}$. Right part of the figure
shows the ideal final setup after the application of LOCC protocol
between ${\cal A}$ and ${\cal R}$. In the ideal case the recipient
${\cal R}$ obtains the full information which was initially present
in the apparatus ${\cal A}$.}
\end{figure}

In recent years, various applications for quantum discord and related
quantum correlations have been discovered. They range from interpreting
the difference between quantum and classical Maxwell's demons \cite{Zurek2003},
to the creation of entanglement in the measurement process \cite{Streltsov2011,Piani2011},
and the use of entanglement in the task of quantum state merging \cite{Madhok2011,Cavalcanti2011}.
Alternative thermodynamical approach has also been presented in \cite{Oppenheim2002}.
Recent results also provide evidence that quantum discord is a resource
in the tasks of entanglement distribution \cite{Streltsov2012,Chuan2012},
remote state preparation \cite{Dakic2012}, and information encoding
\cite{Gu2012}. The role of quantum discord in the historical debate
between EPR and Bohr was also subjected to scrutiny, showing that
quantum discord is closely related to Bohr's criterion for disturbance
\cite{Wiseman2013}. Experimentally friendly measures of quantum discord
have also been considered in \cite{Girolami2012,Silva2013}. Moreover,
quantum discord was identified as the resource for the quantum computing
protocol known as DQC1 \cite{Knill1998,Datta2008}. Remarkably, the
algorithm does not require any entanglement, and it was shown in \cite{Datta2008}
that a typical instance of DQC1 has nonzero quantum discord. 

We revisit Bohr's original idea that a quantum measurement outcome
is classical when it can be communicated by classical means in the
light of recent information-theoretic results. We consider the scenario
illustrated in Fig. \ref{fig:LOCC}. Initially, information about
the system ${\cal S}$ manifests itself in a joint quantum state between
${\cal S}$ and a measurement apparatus ${\cal A}$ (left part of
Fig. \ref{fig:LOCC}). Suppose that the state of the apparatus is
communicated classically to the recipient ${\cal R}$. As is customary
in the quantum information theory, we allow classical communication
and arbitrary quantum operations to be performed locally on ${\cal A}$
and ${\cal R}$. The total procedure is known as ``local operations
and classical communication'' (LOCC) \cite{Bennett1996}. The aim
of this process is to give the recipient ${\cal R}$ all the information
about the system ${\cal S}$, which was initially present in the apparatus
${\cal A}$ (right part of Fig. \ref{fig:LOCC}). If this procedure
was possible for some LOCC protocol, we would say that the measurement
results can be perfectly communicated by classical means. Then, according
to Bohr's postulate, the apparatus would carry solely classical information
about the system ${\cal S}$. If, on the other hand, the process is
not possible for any LOCC protocol, some part of the information about
the system is unavoidably lost on the way to the recipient. In this
case, the apparatus must have contained information about the system,
which could not be communicated by classical means, and thus (according
to Bohr) must have been quantum.

Note that the final state between the system ${\cal S}$ and the recipient
${\cal R}$ is never entangled. This means that the procedure described
above cannot be implemented perfectly when the system is initially
entangled with the apparatus. This observation underscores the quantum
nature of entanglement: all entangled states contain information that
is nonclassical. One might expect that this quantum feature disappears
for all \emph{separable} system-apparatus states, since these states
can be produced using solely classical means. However, this is not
the case: we will see below that even separable states can contain
information which cannot be communicated classically.

In the following, we quantify the information in the apparatus ${\cal A}$
about the system ${\cal S}$ by the quantum mutual information $I({\cal S}:{\cal A})$.
The information gained by the recipient ${\cal R}$ about the system
${\cal S}$ after applying the LOCC protocol is given by their mutual
information $I({\cal S}:{\cal R})$. Our main question can then be
stated as follows: \emph{How much information can a recipient gain
about a system by classical means?} That is, we are interested in
the maximal mutual information between ${\cal S}$ and ${\cal R}$,
maximized over all possible LOCC protocols between the apparatus ${\cal A}$
and the recipient ${\cal R}$. The corresponding quantity will be
called $I^{c}$, where the superscript $c$ tells us that the communication
is classical. Our main result is the following closed expression for
$I^{c}$: 
\begin{equation}
I^{c}=I({\cal S}:{\cal A})-\delta({\cal S}:{\cal A}),\label{eq:Imax}
\end{equation}
which is also equal to the measure of classical correlations introduced
in \cite{Henderson2001}.

The key idea behind the proof of this result is the fact, that the
final state $\rho_{\mathrm{final}}^{{\cal SR}}$ is never entangled,
and that the total initial state has the product form $\rho^{{\cal SA}}\otimes\rho^{{\cal R}}$
\footnote{We note that for general initial states $\rho^{{\cal SAR}}$ it is
possible to create entanglement between ${\cal S}$ and ${\cal R}$
via LOCC between ${\cal A}$ and ${\cal R}$. This can be demonstrated
using the total initial state $\rho=\frac{1}{2}\ket{0}\bra{0}^{{\cal A}}\otimes\sigma^{{\cal SR}}+\frac{1}{2}\ket{1}\bra{1}^{{\cal A}}\otimes\tau^{{\cal SR}}$,
which can have arbitrary little entanglement between ${\cal S}$ and
${\cal R}$, even for highly entangled states $\sigma$ and $\tau$.
By LOCC between ${\cal A}$ and ${\cal R}$ it is possible to create
the final state $\rho'=\frac{1}{2}\ket{0}\bra{0}^{{\cal R}'}\otimes\sigma^{{\cal SR}}+\frac{1}{2}\ket{1}\bra{1}^{{\cal R}'}\otimes\tau^{{\cal SR}}$,
where the recipient is now in possession of ${\cal R}$ and ${\cal R}'$.
Here, the entanglement between the system and the recipient can be
large, even if they had little or even zero initial entanglement.%
}. If we consider the map $\Lambda$ from the initial state $\rho^{{\cal SA}}$
onto the separable final state $\rho_{\mathrm{final}}^{{\cal SR}}=\Lambda(\rho^{{\cal SA}})$,
this map must be entanglement breaking \cite{Horodecki2003}. This
implies that the final state has the form $\rho_{\mathrm{final}}^{{\cal SR}}=\sum_{i}\mathrm{Tr}_{{\cal A}}[E_{i}^{{\cal A}}\rho^{{\cal SA}}]\otimes\sigma_{i}^{{\cal R}}$.
From this result the structure of the optimal LOCC protocol becomes
evident: the apparatus ${\cal A}$ is measured with a measurement
$\left\{ E_{i}^{{\cal A}}\right\} $, and the outcome $i$ is communicated
to the recipient ${\cal R}$, who prepares the state $\sigma_{i}^{{\cal R}}$
locally. The best choice for states $\sigma_{i}^{{\cal R}}$ is to
take them pure and orthogonal \cite{Vedral2002}. In this case the
mutual information becomes $I({\cal S}:{\cal R})=J({\cal S}:{\cal A})_{\left\{ E_{i}^{{\cal A}}\right\} }$.
Eq. (\ref{eq:Imax}) is obtained by taking the maximum over all measurements. 

When the quantum discord between the system ${\cal S}$ and the apparatus
${\cal A}$ is nonzero, $I^{c}$ will always be below the initial
mutual information $I({\cal S}:{\cal A})$. This tells us that \emph{any}
state with nonzero quantum discord contains nonclassical information,
i.e., information which cannot be communicated by classical means.
Interestingly, this statement also includes system-apparatus states
that are separable, i.e., not entangled. This follows from the fact
that quantum discord can be nonzero even when the system is not entangled
with the apparatus \cite{Ollivier2001}. 

The results presented so far fully support Bohr's seminal idea: a
measurement outcome can be regarded as classical when it can be classically
communicated to a distant recipient. From this point of view, the
distinction between the classical and quantum world arises from constraints
on the way information is communicated. We could also paraphrase Wheeler's
summary of Bohr's views and say that ``No phenomenon is a phenomenon
until it is a classically communicable phenomenon'' \cite[p. 182ff.]{Wheeler1983}.
On the other hand, when \emph{quantum communication} between the apparatus
${\cal A}$ and the recipient ${\cal R}$ is allowed, the process
always succeeds, even if the measurement apparatus is nonclassical
\cite{Bennett1993}. Thus, our main result in Eq. (\ref{eq:Imax})
provides a physical interpretation for quantum discord: it measures
the advantage of quantum communication for passing on information.
In particular, this advantage is maximal for maximally entangled states,
since these states also have maximal quantum discord. One might expect
that the advantage remains at least to some degree if the information
is communicated to more than one recipient. However, as we show in
the following, already for two recipients the superiority of quantum
communication disappears for all pure system-apparatus states.

\begin{figure}
\begin{centering}
\includegraphics[width=1\columnwidth]{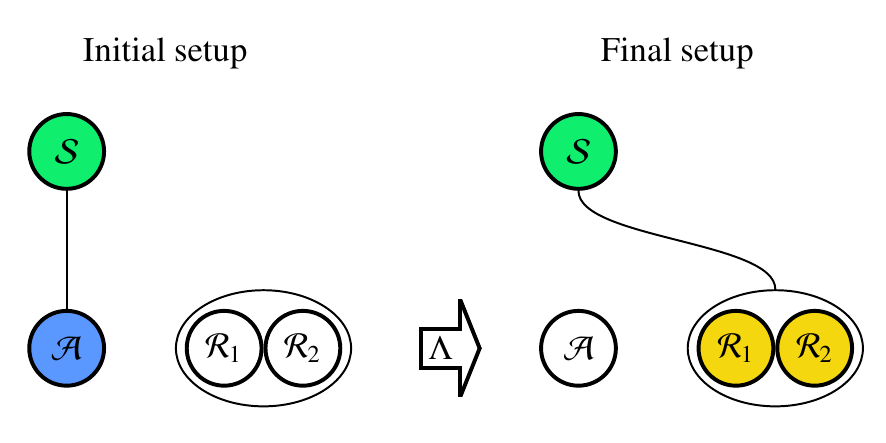} 
\par\end{centering}

\caption{\label{fig:2receivers} Distributing information to two recipients.
Left part of the figure shows the initial situation: The apparatus
${\cal A}$ is initially correlated with the system ${\cal S}$, while
two recipients ${\cal R}_{1}$ and ${\cal R}_{2}$ are not correlated
with ${\cal SA}$. The final situation after the application of a
general quantum channel $\Lambda$ is shown in the right part of the
figure. In the desired final case both recipients have the same information
about the system, $I({\cal S}:{\cal R}_{1})=I({\cal S}:{\cal R}_{2})$.}
\end{figure}
 We consider the system ${\cal S}$ and the apparatus ${\cal A}$,
initially in a joint state (left part of Fig. \ref{fig:2receivers}).
Instead of one recipient ${\cal R}$ we now introduce two recipients
${\cal R}_{1}$ and ${\cal R}_{2}$. The recipients have access to
the apparatus ${\cal A}$ via a communication channel $\Lambda$,
that will be specified below. Aim of the process is to give both recipients
the same amount of information about the system, i.e., $I({\cal S}:{\cal R}_{1})=I({\cal S}:{\cal R}_{2})$
(right part of Fig. \ref{fig:2receivers}). Applying the same line
of reasoning as before, we arrive at the following question: \emph{How
much information can each of two recipients gain about a system? }We
denote the maximal amount of such shared information attainable via
LOCC by $I_{2}^{c}$, where the lower index gives the number of recipients.
From the previous results is clear that the number of recipients does
not change anything in the case of classical communication: $I_{n}^{c}=I^{c}$
for any number of recipients $n$.

We will now compare this quantity to the amount of mutual information
attainable via \emph{quantum communication}, which will be denoted
by $I_{2}^{q}$. Since quantum communication is more general than
any LOCC protocol, it must be that $I_{n}^{q}\geq I_{n}^{c}$, which
is indeed true for any number of recipients $n$. In the single-recipient
scenario this inequality becomes strict for all states with nonzero
quantum discord: in all these cases quantum communication provides
an advantage. In particular, for all pure system-apparatus states
we have $I_{1}^{q}=2I_{1}^{c}$, i.e., quantum communication can outperform
the best LOCC protocol by a factor of $2$. However, as we show in
the following, quantum communication does not provide any advantage
for any number of recipients $n\geq2$, if the initial system-apparatus
state is pure: 
\begin{equation}
I_{n}^{q}=I_{n}^{c}.\label{eq:nreceivers}
\end{equation}

For proving this result we first consider the two-recipients scenario,
i.e., $n=2$. The amount of information attainable by classical means
is $I_{2}^{c}=S(\rho^{{\cal S}})$. We will now show that $I_{2}^{q}=I_{2}^{c}$.
For this we can assume that each recipient initially has the state
$\ket{0}^{{\cal R}_{i}}$, i.e., the total initial state is $\ket{\psi_{\mathrm{tot}}}=\ket{\psi}^{{\cal SA}}\ket{0}^{{\cal R}_{1}}\ket{0}^{{\cal R}_{2}}$.
After the application of a quantum channel, the final state can be
written as $\rho_{\mathrm{final}}^{{\cal SAR}_{1}{\cal R}_{2}}=\mathrm{Tr}_{{\cal B}}[U\rho_{\mathrm{tot}}U^{\dagger}]$,
where ${\cal B}$ is an ancilla, $\rho_{\mathrm{tot}}=\ket{\psi_{\mathrm{tot}}}\bra{\psi_{\mathrm{tot}}}\otimes\ket{0}\bra{0}^{{\cal B}}$,
and $U$ is a unitary acting on ${\cal AR}_{1}{\cal R}_{2}{\cal B}$.
Since mutual information does not increase under partial trace \cite{Vedral2002},
it follows that $I({\cal S}:{\cal R}_{1})+I({\cal S}:{\cal R}_{2})\leq I({\cal S}:{\cal AR}_{1})+I({\cal S}:{\cal R}_{2}{\cal B})$,
where $\rho^{{\cal SR}_{2}{\cal B}}=\mathrm{Tr}_{{\cal AR}_{1}}[U\rho_{\mathrm{tot}}U^{\dagger}]$.
On the other hand, since the total system ${\cal SAR}_{1}{\cal R}_{2}{\cal B}$
is in a pure state, we get $S(\rho^{{\cal AR}_{1}})=S(\rho^{{\cal SR}_{2}{\cal B}})$
and $S(\rho^{{\cal R}_{2}{\cal B}})=S(\rho^{{\cal SAR}_{1}})$ so
that $I({\cal S}:{\cal AR}_{1})+I({\cal S}:{\cal R}_{2}{\cal B})=2S(\rho^{{\cal S}})$,
leading to the inequality $I({\cal S}:{\cal R}_{1})+I({\cal S}:{\cal R}_{2})\leq2I_{2}^{c}$.
Since we demand that $I({\cal S}:{\cal R}_{1})=I({\cal S}:{\cal R}_{2})$,
both quantities cannot be larger than $I_{2}^{c}$, which implies
$I_{2}^{q}=I_{2}^{c}$. Note that this reasoning also implies the
equality $I_{n}^{q}=I_{n}^{c}$ for any number of recipients $n\geq2$,
since it shows that for each pair of recipients ${\cal R}_{i}$ and
${\cal R}_{j}$ with $i\neq j$ the sum $I({\cal S}:{\cal R}_{i})+I({\cal S}:{\cal R}_{j})$
never exceeds $2I_{2}^{c}$. These results can also be generalized
to asymmetric protocols, where the final mutual information between
the system and each of $n$ recipients is not necessarily the same.
In this case, quantum communication cannot outperform LOCC \emph{on
average}: $\frac{1}{n}\sum_{i=1}^{n}I({\cal S}:{\cal R}_{i})\leq I_{n}^{c}$.
This can be seen using the same arguments as above, noting that the
sum $\sum_{i=1}^{n}I({\cal S}:{\cal R}_{i})$ never exceeds $n\cdot I_{n}^{c}$.

\begin{figure}
\begin{centering}
\includegraphics[width=1\columnwidth]{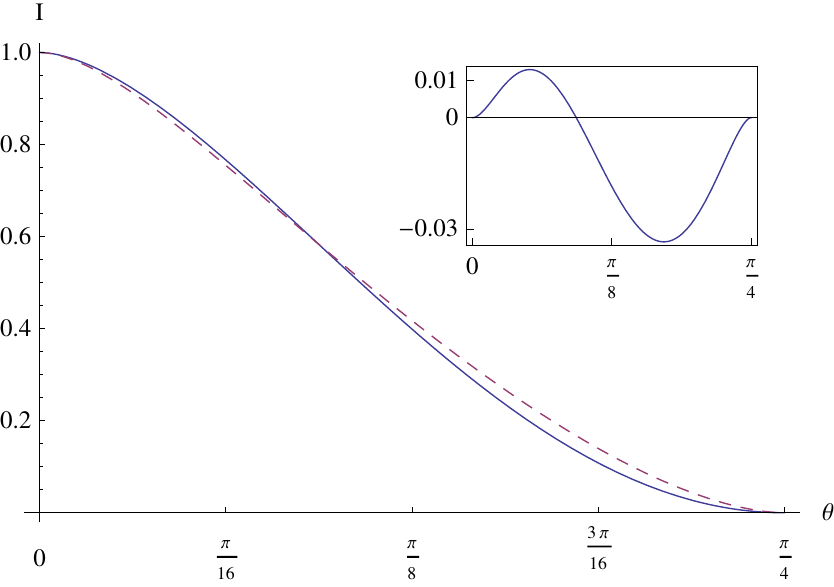}
\par\end{centering}

\caption{\label{fig:plot} Main figure shows the maximal amount of information
attainable by classical means $I^{c}$ (solid line) and the mutual
information $I({\cal S}:{\cal R}_{1})=I({\cal S}:{\cal R}_{2})$ for
the ``optimal state-dependent cloning'' (dashed line) for the state
given in Eq. (\ref{eq:rho}) as a function of $\theta$. The inset
shows the difference $I^{c}-I({\cal S}:{\cal R}_{1})$. For $\theta'<\theta<\pi/4$
with $\theta'\approx0.093\pi$ quantum communication outperforms any
LOCC protocol, see main text for details.}
\end{figure}

It is now natural to ask whether Eq. (\ref{eq:nreceivers}) also holds
for initially mixed states $\rho^{{\cal SA}}$. We will see in the
following that in general this is not the case: for mixed system-apparatus
states quantum communication can outperform classical communication.
This will be demonstrated using the initial state 
\begin{equation}
\rho^{{\cal SA}}=\frac{1}{2}\ket{0}\bra{0}^{{\cal S}}\otimes\ket{\psi}\bra{\psi}^{{\cal A}}+\frac{1}{2}\ket{1}\bra{1}^{{\cal S}}\otimes\ket{\phi}\bra{\phi}^{{\cal A}}\label{eq:rho}
\end{equation}
 with $\ket{\psi}=\cos\theta\ket{0}+\sin\theta\ket{1}$ and $\ket{\phi}=\sin\theta\ket{0}+\cos\theta\ket{1}$.
For this initial state, $I^{c}$ can be evaluated using the Koashi-Winter
relation \cite{Koashi2004}. The result is shown in Fig. \ref{fig:plot}. 

On the other hand, the maximal amount of information $I_{2}^{q}$
attainable via a quantum channel can be bounded below by any imperfect
cloning protocol for the two states $\ket{\psi}$ and $\ket{\phi}$.$ $
In the following we will use the protocol for ``optimal state-dependent
cloning'' presented in \cite{Bruss1998}. The corresponding mutual
information $I({\cal S}:{\cal R}_{1})=I({\cal S}:{\cal R}_{2})$ between
the system ${\cal S}$ and each of the recipients ${\cal R}_{1}$
and ${\cal R}_{2}$ is shown in Fig. \ref{fig:plot}. As can be seen
from the difference $I^{c}-I({\cal S}:{\cal R}_{1})$ shown in the
inset of Fig. \ref{fig:plot}, the cloning procedure outperforms any
LOCC protocol in the region $\theta'<\theta<\pi/4$ with $\theta'\approx0.093\pi$.
However, quite surprisingly, we also see that LOCC outperforms the
protocol of optimal state-dependent cloning for $0<\theta<\theta'$.
The reason for this counterintuitive behavior is the fact that the
protocol considered in \cite{Bruss1998} was optimized for a different
figure of merit known as global fidelity. These results demonstrate
the fundamental difference between the new task of distributing correlations
considered in this Letter and the task of cloning a quantum state
\cite{Wootters1982,Dieks1982}. 

The results presented so far can be seen as an extension of Bohr's
postulate: classicality emerges also whenever the outcome of a measurement
is shared by more than one recipient. As sharing of information is
a prerequisite for ``objective reality'' \cite{Zurek2009,Riedel2010,Zwolak2013},
our results indicate that objective information is necessarily classical:
For pure system-apparatus states classicality arises whenever data
are communicated to more than one party, independently from the nature
of the communication channel. Moreover, a natural operational definition
of discord turns out to be a counterpoint of the operational definition
of entanglement: A composite state that cannot be assembled by classical
means is entangled. Similarly, a composite state that cannot be disassembled
by classical means is still quantum -- it has a nonvanishing discord,
quantum information lost in the process of deconstructing it into
classical ingredients.

Finally, we point out that the single-recipient scenario, as shown
in Fig. \ref{fig:LOCC}, can also be regarded as a modified version
of the quantum state merging \cite{Horodecki2005}. There are however
two essential differences to the situation considered in \cite{Horodecki2005}.
We do not demand that the total system ${\cal SAR}$ is in a pure
state and, in contrast to \cite{Horodecki2005}, we do not allow any
entanglement to be shared between ${\cal A}$ and ${\cal R}$. Moreover,
the discussion in \cite{Horodecki2005} concentrates on the scenario
where many copies of the same state are available. We expect that
a fruitful comparison of the two approaches can be made, if the results
of our work are also extended to the many-copy case. On the other
hand, the situation of many recipients, as illustrated in Fig. \ref{fig:2receivers},
can be related to the protocol known as local broadcasting \cite{Piani2008,Luo2010}.
While in Ref. \cite{Luo2010} the authors conclude that perfect local
broadcasting is only possible for states with zero quantum discord,
our result in Eq. (\ref{eq:nreceivers}) can be regarded as a generalization
to the case of all pure states. From this point of view, our results
imply that local broadcasting of a pure state never requires quantum
communication; the best performance is always achievable with LOCC. 

We conclude with Bohr's statement \cite{Bohr1963}: \emph{``No observation
is an observation unless we can communicate the results of that observation
to others in plain language.''} In this work we have shown that this
statement is not only philosophical: if measurement outcomes are to
be communicated to distant recipients by classical means, the measurement
apparatus must be in a classical state. In this sense, classicality
naturally arises when communication is restricted to plain language.
Moreover, we have applied our results to the case when measurement
outcomes are communicated to more than one recipient. We found that
for \emph{pure} system-apparatus states quantum communication does
not provide any advantage, showing that classicality arises in this
case regardless of the nature of the communication. Although quantum
communication is superior to classical communication on \emph{mixed}
system-apparatus states, there is strong evidence \cite{Bae2006,Chiribella2006,Brandao2013}
that in the limit $n\rightarrow\infty$, i.e., for a large number
of recipients, the result in Eq. (\ref{eq:nreceivers}) becomes valid
for all mixed states. A rigorous proof of this statement is left open
for future research.

We acknowledge discussion with Gerardo Adesso, Dagmar Bruß, Otfried
Gühne, Hermann Kampermann, Matthias Kleinmann, Tobias Moroder, Marco
Piani, Volkher Scholz, and Michael Zwolak. This work was supported
by DoE LDRD program at Los Alamos, and in part by John Templeton Foundation,
Deutsche Forschungsgemeinschaft (DFG) and ELES\texttt{.}

 \bibliographystyle{apsrev4-1}
\bibliography{literature}

\begin{thebibliography}{47}%
\makeatletter
\providecommand \@ifxundefined [1]{%
 \@ifx{#1\undefined}
}%
\providecommand \@ifnum [1]{%
 \ifnum #1\expandafter \@firstoftwo
 \else \expandafter \@secondoftwo
 \fi
}%
\providecommand \@ifx [1]{%
 \ifx #1\expandafter \@firstoftwo
 \else \expandafter \@secondoftwo
 \fi
}%
\providecommand \natexlab [1]{#1}%
\providecommand \enquote  [1]{``#1''}%
\providecommand \bibnamefont  [1]{#1}%
\providecommand \bibfnamefont [1]{#1}%
\providecommand \citenamefont [1]{#1}%
\providecommand \href@noop [0]{\@secondoftwo}%
\providecommand \href [0]{\begingroup \@sanitize@url \@href}%
\providecommand \@href[1]{\@@startlink{#1}\@@href}%
\providecommand \@@href[1]{\endgroup#1\@@endlink}%
\providecommand \@sanitize@url [0]{\catcode `\\12\catcode `\$12\catcode
  `\&12\catcode `\#12\catcode `\^12\catcode `\_12\catcode `\%12\relax}%
\providecommand \@@startlink[1]{}%
\providecommand \@@endlink[0]{}%
\providecommand \url  [0]{\begingroup\@sanitize@url \@url }%
\providecommand \@url [1]{\endgroup\@href {#1}{\urlprefix }}%
\providecommand \urlprefix  [0]{URL }%
\providecommand \Eprint [0]{\href }%
\providecommand \doibase [0]{http://dx.doi.org/}%
\providecommand \selectlanguage [0]{\@gobble}%
\providecommand \bibinfo  [0]{\@secondoftwo}%
\providecommand \bibfield  [0]{\@secondoftwo}%
\providecommand \translation [1]{[#1]}%
\providecommand \BibitemOpen [0]{}%
\providecommand \bibitemStop [0]{}%
\providecommand \bibitemNoStop [0]{.\EOS\space}%
\providecommand \EOS [0]{\spacefactor3000\relax}%
\providecommand \BibitemShut  [1]{\csname bibitem#1\endcsname}%
\let\auto@bib@innerbib\@empty
\bibitem [{\citenamefont {Schr\"{o}dinger}(1935)}]{Schroedinger1935}%
  \BibitemOpen
  \bibfield  {author} {\bibinfo {author} {\bibfnamefont {E.}~\bibnamefont
  {Schr\"{o}dinger}},\ }\href {\doibase 10.1007/BF01491891} {\bibfield
  {journal} {\bibinfo  {journal} {Naturwissenschaften}\ }\textbf {\bibinfo
  {volume} {23}},\ \bibinfo {pages} {807} (\bibinfo {year} {1935})}\BibitemShut
  {NoStop}%
\bibitem [{\citenamefont {Zurek}(1981)}]{Zurek1981}%
  \BibitemOpen
  \bibfield  {author} {\bibinfo {author} {\bibfnamefont {W.~H.}\ \bibnamefont
  {Zurek}},\ }\href {\doibase 10.1103/PhysRevD.24.1516} {\bibfield  {journal}
  {\bibinfo  {journal} {Phys. Rev. D}\ }\textbf {\bibinfo {volume} {24}},\
  \bibinfo {pages} {1516} (\bibinfo {year} {1981})}\BibitemShut {NoStop}%
\bibitem [{\citenamefont {Einstein}\ \emph {et~al.}(1935)\citenamefont
  {Einstein}, \citenamefont {Podolsky},\ and\ \citenamefont
  {Rosen}}]{Einstein1935}%
  \BibitemOpen
  \bibfield  {author} {\bibinfo {author} {\bibfnamefont {A.}~\bibnamefont
  {Einstein}}, \bibinfo {author} {\bibfnamefont {B.}~\bibnamefont {Podolsky}},
  \ and\ \bibinfo {author} {\bibfnamefont {N.}~\bibnamefont {Rosen}},\ }\href
  {\doibase 10.1103/PhysRev.47.777} {\bibfield  {journal} {\bibinfo  {journal}
  {Phys. Rev.}\ }\textbf {\bibinfo {volume} {47}},\ \bibinfo {pages} {777}
  (\bibinfo {year} {1935})}\BibitemShut {NoStop}%
\bibitem [{\citenamefont {Zurek}(2003{\natexlab{a}})}]{Zurek2003a}%
  \BibitemOpen
  \bibfield  {author} {\bibinfo {author} {\bibfnamefont {W.~H.}\ \bibnamefont
  {Zurek}},\ }\href {\doibase 10.1103/PhysRevLett.90.120404} {\bibfield
  {journal} {\bibinfo  {journal} {Phys. Rev. Lett.}\ }\textbf {\bibinfo
  {volume} {90}},\ \bibinfo {pages} {120404} (\bibinfo {year}
  {2003}{\natexlab{a}})}\BibitemShut {NoStop}%
\bibitem [{\citenamefont {Zurek}(2005)}]{Zurek2005}%
  \BibitemOpen
  \bibfield  {author} {\bibinfo {author} {\bibfnamefont {W.~H.}\ \bibnamefont
  {Zurek}},\ }\href {\doibase 10.1103/PhysRevA.71.052105} {\bibfield  {journal}
  {\bibinfo  {journal} {Phys. Rev. A}\ }\textbf {\bibinfo {volume} {71}},\
  \bibinfo {pages} {052105} (\bibinfo {year} {2005})}\BibitemShut {NoStop}%
\bibitem [{\citenamefont {Zurek}(2011)}]{Zurek2011}%
  \BibitemOpen
  \bibfield  {author} {\bibinfo {author} {\bibfnamefont {W.~H.}\ \bibnamefont
  {Zurek}},\ }\href {\doibase 10.1103/PhysRevLett.106.250402} {\bibfield
  {journal} {\bibinfo  {journal} {Phys. Rev. Lett.}\ }\textbf {\bibinfo
  {volume} {106}},\ \bibinfo {pages} {250402} (\bibinfo {year}
  {2011})}\BibitemShut {NoStop}%
\bibitem [{\citenamefont {Werner}(1989)}]{Werner1989}%
  \BibitemOpen
  \bibfield  {author} {\bibinfo {author} {\bibfnamefont {R.~F.}\ \bibnamefont
  {Werner}},\ }\href {\doibase 10.1103/PhysRevA.40.4277} {\bibfield  {journal}
  {\bibinfo  {journal} {Phys. Rev. A}\ }\textbf {\bibinfo {volume} {40}},\
  \bibinfo {pages} {4277} (\bibinfo {year} {1989})}\BibitemShut {NoStop}%
\bibitem [{\citenamefont {Cover}\ and\ \citenamefont
  {Thomas}(1991)}]{Cover1991}%
  \BibitemOpen
  \bibfield  {author} {\bibinfo {author} {\bibfnamefont {T.~M.}\ \bibnamefont
  {Cover}}\ and\ \bibinfo {author} {\bibfnamefont {J.~A.}\ \bibnamefont
  {Thomas}},\ }\href@noop {} {\emph {\bibinfo {title} {Elements of Information
  Theory}}}\ (\bibinfo  {publisher} {J. Wiley},\ \bibinfo {address} {New
  York},\ \bibinfo {year} {1991})\BibitemShut {NoStop}%
\bibitem [{\citenamefont {Zurek}(2000)}]{Zurek2000}%
  \BibitemOpen
  \bibfield  {author} {\bibinfo {author} {\bibfnamefont {W.~H.}\ \bibnamefont
  {Zurek}},\ }\href {\doibase
  10.1002/1521-3889(200011)9:11/12<855::AID-ANDP855>3.0.CO;2-K} {\bibfield
  {journal} {\bibinfo  {journal} {Ann. Phys. (Berlin)}\ }\textbf {\bibinfo
  {volume} {9}},\ \bibinfo {pages} {855} (\bibinfo {year} {2000})}\BibitemShut
  {NoStop}%
\bibitem [{\citenamefont {Ollivier}\ and\ \citenamefont
  {Zurek}(2001)}]{Ollivier2001}%
  \BibitemOpen
  \bibfield  {author} {\bibinfo {author} {\bibfnamefont {H.}~\bibnamefont
  {Ollivier}}\ and\ \bibinfo {author} {\bibfnamefont {W.~H.}\ \bibnamefont
  {Zurek}},\ }\href {\doibase 10.1103/PhysRevLett.88.017901} {\bibfield
  {journal} {\bibinfo  {journal} {Phys. Rev. Lett.}\ }\textbf {\bibinfo
  {volume} {88}},\ \bibinfo {pages} {017901} (\bibinfo {year}
  {2001})}\BibitemShut {NoStop}%
\bibitem [{\citenamefont {Henderson}\ and\ \citenamefont
  {Vedral}(2001)}]{Henderson2001}%
  \BibitemOpen
  \bibfield  {author} {\bibinfo {author} {\bibfnamefont {L.}~\bibnamefont
  {Henderson}}\ and\ \bibinfo {author} {\bibfnamefont {V.}~\bibnamefont
  {Vedral}},\ }\href {\doibase 10.1088/0305-4470/34/35/315} {\bibfield
  {journal} {\bibinfo  {journal} {J. Phys. A}\ }\textbf {\bibinfo {volume}
  {34}},\ \bibinfo {pages} {6899} (\bibinfo {year} {2001})}\BibitemShut
  {NoStop}%
\bibitem [{\citenamefont {Modi}\ \emph {et~al.}(2012)\citenamefont {Modi},
  \citenamefont {Brodutch}, \citenamefont {Cable}, \citenamefont {Paterek},\
  and\ \citenamefont {Vedral}}]{Modi2012}%
  \BibitemOpen
  \bibfield  {author} {\bibinfo {author} {\bibfnamefont {K.}~\bibnamefont
  {Modi}}, \bibinfo {author} {\bibfnamefont {A.}~\bibnamefont {Brodutch}},
  \bibinfo {author} {\bibfnamefont {H.}~\bibnamefont {Cable}}, \bibinfo
  {author} {\bibfnamefont {T.}~\bibnamefont {Paterek}}, \ and\ \bibinfo
  {author} {\bibfnamefont {V.}~\bibnamefont {Vedral}},\ }\href {\doibase
  10.1103/RevModPhys.84.1655} {\bibfield  {journal} {\bibinfo  {journal} {Rev.
  Mod. Phys.}\ }\textbf {\bibinfo {volume} {84}},\ \bibinfo {pages} {1655}
  (\bibinfo {year} {2012})}\BibitemShut {NoStop}%
\bibitem [{\citenamefont {Zurek}(2003{\natexlab{b}})}]{Zurek2003}%
  \BibitemOpen
  \bibfield  {author} {\bibinfo {author} {\bibfnamefont {W.~H.}\ \bibnamefont
  {Zurek}},\ }\href {\doibase 10.1103/PhysRevA.67.012320} {\bibfield  {journal}
  {\bibinfo  {journal} {Phys. Rev. A}\ }\textbf {\bibinfo {volume} {67}},\
  \bibinfo {pages} {012320} (\bibinfo {year} {2003}{\natexlab{b}})}\BibitemShut
  {NoStop}%
\bibitem [{\citenamefont {Streltsov}\ \emph {et~al.}(2011)\citenamefont
  {Streltsov}, \citenamefont {Kampermann},\ and\ \citenamefont
  {Bru\ss{}}}]{Streltsov2011}%
  \BibitemOpen
  \bibfield  {author} {\bibinfo {author} {\bibfnamefont {A.}~\bibnamefont
  {Streltsov}}, \bibinfo {author} {\bibfnamefont {H.}~\bibnamefont
  {Kampermann}}, \ and\ \bibinfo {author} {\bibfnamefont {D.}~\bibnamefont
  {Bru\ss{}}},\ }\href {\doibase 10.1103/PhysRevLett.106.160401} {\bibfield
  {journal} {\bibinfo  {journal} {Phys. Rev. Lett.}\ }\textbf {\bibinfo
  {volume} {106}},\ \bibinfo {pages} {160401} (\bibinfo {year}
  {2011})}\BibitemShut {NoStop}%
\bibitem [{\citenamefont {Piani}\ \emph {et~al.}(2011)\citenamefont {Piani},
  \citenamefont {Gharibian}, \citenamefont {Adesso}, \citenamefont
  {Calsamiglia}, \citenamefont {Horodecki},\ and\ \citenamefont
  {Winter}}]{Piani2011}%
  \BibitemOpen
  \bibfield  {author} {\bibinfo {author} {\bibfnamefont {M.}~\bibnamefont
  {Piani}}, \bibinfo {author} {\bibfnamefont {S.}~\bibnamefont {Gharibian}},
  \bibinfo {author} {\bibfnamefont {G.}~\bibnamefont {Adesso}}, \bibinfo
  {author} {\bibfnamefont {J.}~\bibnamefont {Calsamiglia}}, \bibinfo {author}
  {\bibfnamefont {P.}~\bibnamefont {Horodecki}}, \ and\ \bibinfo {author}
  {\bibfnamefont {A.}~\bibnamefont {Winter}},\ }\href {\doibase
  10.1103/PhysRevLett.106.220403} {\bibfield  {journal} {\bibinfo  {journal}
  {Phys. Rev. Lett.}\ }\textbf {\bibinfo {volume} {106}},\ \bibinfo {pages}
  {220403} (\bibinfo {year} {2011})}\BibitemShut {NoStop}%
\bibitem [{\citenamefont {Madhok}\ and\ \citenamefont
  {Datta}(2011)}]{Madhok2011}%
  \BibitemOpen
  \bibfield  {author} {\bibinfo {author} {\bibfnamefont {V.}~\bibnamefont
  {Madhok}}\ and\ \bibinfo {author} {\bibfnamefont {A.}~\bibnamefont {Datta}},\
  }\href {\doibase 10.1103/PhysRevA.83.032323} {\bibfield  {journal} {\bibinfo
  {journal} {Phys. Rev. A}\ }\textbf {\bibinfo {volume} {83}},\ \bibinfo
  {pages} {032323} (\bibinfo {year} {2011})}\BibitemShut {NoStop}%
\bibitem [{\citenamefont {Cavalcanti}\ \emph {et~al.}(2011)\citenamefont
  {Cavalcanti}, \citenamefont {Aolita}, \citenamefont {Boixo}, \citenamefont
  {Modi}, \citenamefont {Piani},\ and\ \citenamefont
  {Winter}}]{Cavalcanti2011}%
  \BibitemOpen
  \bibfield  {author} {\bibinfo {author} {\bibfnamefont {D.}~\bibnamefont
  {Cavalcanti}}, \bibinfo {author} {\bibfnamefont {L.}~\bibnamefont {Aolita}},
  \bibinfo {author} {\bibfnamefont {S.}~\bibnamefont {Boixo}}, \bibinfo
  {author} {\bibfnamefont {K.}~\bibnamefont {Modi}}, \bibinfo {author}
  {\bibfnamefont {M.}~\bibnamefont {Piani}}, \ and\ \bibinfo {author}
  {\bibfnamefont {A.}~\bibnamefont {Winter}},\ }\href {\doibase
  10.1103/PhysRevA.83.032324} {\bibfield  {journal} {\bibinfo  {journal} {Phys.
  Rev. A}\ }\textbf {\bibinfo {volume} {83}},\ \bibinfo {pages} {032324}
  (\bibinfo {year} {2011})}\BibitemShut {NoStop}%
\bibitem [{\citenamefont {Oppenheim}\ \emph {et~al.}(2002)\citenamefont
  {Oppenheim}, \citenamefont {Horodecki}, \citenamefont {Horodecki},\ and\
  \citenamefont {Horodecki}}]{Oppenheim2002}%
  \BibitemOpen
  \bibfield  {author} {\bibinfo {author} {\bibfnamefont {J.}~\bibnamefont
  {Oppenheim}}, \bibinfo {author} {\bibfnamefont {M.}~\bibnamefont
  {Horodecki}}, \bibinfo {author} {\bibfnamefont {P.}~\bibnamefont
  {Horodecki}}, \ and\ \bibinfo {author} {\bibfnamefont {R.}~\bibnamefont
  {Horodecki}},\ }\href {\doibase 10.1103/PhysRevLett.89.180402} {\bibfield
  {journal} {\bibinfo  {journal} {Phys. Rev. Lett.}\ }\textbf {\bibinfo
  {volume} {89}},\ \bibinfo {pages} {180402} (\bibinfo {year}
  {2002})}\BibitemShut {NoStop}%
\bibitem [{\citenamefont {Streltsov}\ \emph {et~al.}(2012)\citenamefont
  {Streltsov}, \citenamefont {Kampermann},\ and\ \citenamefont
  {Bru\ss{}}}]{Streltsov2012}%
  \BibitemOpen
  \bibfield  {author} {\bibinfo {author} {\bibfnamefont {A.}~\bibnamefont
  {Streltsov}}, \bibinfo {author} {\bibfnamefont {H.}~\bibnamefont
  {Kampermann}}, \ and\ \bibinfo {author} {\bibfnamefont {D.}~\bibnamefont
  {Bru\ss{}}},\ }\href {\doibase 10.1103/PhysRevLett.108.250501} {\bibfield
  {journal} {\bibinfo  {journal} {Phys. Rev. Lett.}\ }\textbf {\bibinfo
  {volume} {108}},\ \bibinfo {pages} {250501} (\bibinfo {year}
  {2012})}\BibitemShut {NoStop}%
\bibitem [{\citenamefont {Chuan}\ \emph {et~al.}(2012)\citenamefont {Chuan},
  \citenamefont {Maillard}, \citenamefont {Modi}, \citenamefont {Paterek},
  \citenamefont {Paternostro},\ and\ \citenamefont {Piani}}]{Chuan2012}%
  \BibitemOpen
  \bibfield  {author} {\bibinfo {author} {\bibfnamefont {T.~K.}\ \bibnamefont
  {Chuan}}, \bibinfo {author} {\bibfnamefont {J.}~\bibnamefont {Maillard}},
  \bibinfo {author} {\bibfnamefont {K.}~\bibnamefont {Modi}}, \bibinfo {author}
  {\bibfnamefont {T.}~\bibnamefont {Paterek}}, \bibinfo {author} {\bibfnamefont
  {M.}~\bibnamefont {Paternostro}}, \ and\ \bibinfo {author} {\bibfnamefont
  {M.}~\bibnamefont {Piani}},\ }\href {\doibase 10.1103/PhysRevLett.109.070501}
  {\bibfield  {journal} {\bibinfo  {journal} {Phys. Rev. Lett.}\ }\textbf
  {\bibinfo {volume} {109}},\ \bibinfo {pages} {070501} (\bibinfo {year}
  {2012})}\BibitemShut {NoStop}%
\bibitem [{\citenamefont {Daki\'{c}}\ \emph {et~al.}(2012)\citenamefont
  {Daki\'{c}}, \citenamefont {Lipp}, \citenamefont {Ma}, \citenamefont
  {Ringbauer}, \citenamefont {Kropatschek}, \citenamefont {Barz}, \citenamefont
  {Paterek}, \citenamefont {Vedral}, \citenamefont {Zeilinger}, \citenamefont
  {Brukner},\ and\ \citenamefont {Walther}}]{Dakic2012}%
  \BibitemOpen
  \bibfield  {author} {\bibinfo {author} {\bibfnamefont {B.}~\bibnamefont
  {Daki\'{c}}}, \bibinfo {author} {\bibfnamefont {Y.~O.}\ \bibnamefont {Lipp}},
  \bibinfo {author} {\bibfnamefont {X.}~\bibnamefont {Ma}}, \bibinfo {author}
  {\bibfnamefont {M.}~\bibnamefont {Ringbauer}}, \bibinfo {author}
  {\bibfnamefont {S.}~\bibnamefont {Kropatschek}}, \bibinfo {author}
  {\bibfnamefont {S.}~\bibnamefont {Barz}}, \bibinfo {author} {\bibfnamefont
  {T.}~\bibnamefont {Paterek}}, \bibinfo {author} {\bibfnamefont
  {V.}~\bibnamefont {Vedral}}, \bibinfo {author} {\bibfnamefont
  {A.}~\bibnamefont {Zeilinger}}, \bibinfo {author} {\bibfnamefont
  {{\v{C}}.}~\bibnamefont {Brukner}}, \ and\ \bibinfo {author} {\bibfnamefont
  {P.}~\bibnamefont {Walther}},\ }\href {\doibase 10.1038/nphys2377} {\bibfield
   {journal} {\bibinfo  {journal} {Nat. Phys.}\ }\textbf {\bibinfo {volume}
  {8}},\ \bibinfo {pages} {666} (\bibinfo {year} {2012})}\BibitemShut {NoStop}%
\bibitem [{\citenamefont {Gu}\ \emph {et~al.}(2012)\citenamefont {Gu},
  \citenamefont {Chrzanowski}, \citenamefont {Assad}, \citenamefont {Symul},
  \citenamefont {Modi}, \citenamefont {Ralph}, \citenamefont {Vedral},\ and\
  \citenamefont {Lam}}]{Gu2012}%
  \BibitemOpen
  \bibfield  {author} {\bibinfo {author} {\bibfnamefont {M.}~\bibnamefont
  {Gu}}, \bibinfo {author} {\bibfnamefont {H.~M.}\ \bibnamefont {Chrzanowski}},
  \bibinfo {author} {\bibfnamefont {S.~M.}\ \bibnamefont {Assad}}, \bibinfo
  {author} {\bibfnamefont {T.}~\bibnamefont {Symul}}, \bibinfo {author}
  {\bibfnamefont {K.}~\bibnamefont {Modi}}, \bibinfo {author} {\bibfnamefont
  {T.~C.}\ \bibnamefont {Ralph}}, \bibinfo {author} {\bibfnamefont
  {V.}~\bibnamefont {Vedral}}, \ and\ \bibinfo {author} {\bibfnamefont {P.~K.}\
  \bibnamefont {Lam}},\ }\href {\doibase 10.1038/nphys2376} {\bibfield
  {journal} {\bibinfo  {journal} {Nat. Phys.}\ }\textbf {\bibinfo {volume}
  {8}},\ \bibinfo {pages} {671} (\bibinfo {year} {2012})}\BibitemShut {NoStop}%
\bibitem [{\citenamefont {Wiseman}()}]{Wiseman2013}%
  \BibitemOpen
  \bibfield  {author} {\bibinfo {author} {\bibfnamefont {H.~M.}\ \bibnamefont
  {Wiseman}},\ }\href@noop {} {}\Eprint {http://arxiv.org/abs/1208.4964v2}
  {arXiv:1208.4964v2} \BibitemShut {NoStop}%
\bibitem [{\citenamefont {Girolami}\ and\ \citenamefont
  {Adesso}(2012)}]{Girolami2012}%
  \BibitemOpen
  \bibfield  {author} {\bibinfo {author} {\bibfnamefont {D.}~\bibnamefont
  {Girolami}}\ and\ \bibinfo {author} {\bibfnamefont {G.}~\bibnamefont
  {Adesso}},\ }\href {\doibase 10.1103/PhysRevLett.108.150403} {\bibfield
  {journal} {\bibinfo  {journal} {Phys. Rev. Lett.}\ }\textbf {\bibinfo
  {volume} {108}},\ \bibinfo {pages} {150403} (\bibinfo {year}
  {2012})}\BibitemShut {NoStop}%
\bibitem [{\citenamefont {Silva}\ \emph {et~al.}(2013)\citenamefont {Silva},
  \citenamefont {Girolami}, \citenamefont {Auccaise}, \citenamefont {Sarthour},
  \citenamefont {Oliveira}, \citenamefont {Bonagamba}, \citenamefont
  {deAzevedo}, \citenamefont {Soares-Pinto},\ and\ \citenamefont
  {Adesso}}]{Silva2013}%
  \BibitemOpen
  \bibfield  {author} {\bibinfo {author} {\bibfnamefont {I.~A.}\ \bibnamefont
  {Silva}}, \bibinfo {author} {\bibfnamefont {D.}~\bibnamefont {Girolami}},
  \bibinfo {author} {\bibfnamefont {R.}~\bibnamefont {Auccaise}}, \bibinfo
  {author} {\bibfnamefont {R.~S.}\ \bibnamefont {Sarthour}}, \bibinfo {author}
  {\bibfnamefont {I.~S.}\ \bibnamefont {Oliveira}}, \bibinfo {author}
  {\bibfnamefont {T.~J.}\ \bibnamefont {Bonagamba}}, \bibinfo {author}
  {\bibfnamefont {E.~R.}\ \bibnamefont {deAzevedo}}, \bibinfo {author}
  {\bibfnamefont {D.~O.}\ \bibnamefont {Soares-Pinto}}, \ and\ \bibinfo
  {author} {\bibfnamefont {G.}~\bibnamefont {Adesso}},\ }\href {\doibase
  10.1103/PhysRevLett.110.140501} {\bibfield  {journal} {\bibinfo  {journal}
  {Phys. Rev. Lett.}\ }\textbf {\bibinfo {volume} {110}},\ \bibinfo {pages}
  {140501} (\bibinfo {year} {2013})}\BibitemShut {NoStop}%
\bibitem [{\citenamefont {Knill}\ and\ \citenamefont
  {Laflamme}(1998)}]{Knill1998}%
  \BibitemOpen
  \bibfield  {author} {\bibinfo {author} {\bibfnamefont {E.}~\bibnamefont
  {Knill}}\ and\ \bibinfo {author} {\bibfnamefont {R.}~\bibnamefont
  {Laflamme}},\ }\href {\doibase 10.1103/PhysRevLett.81.5672} {\bibfield
  {journal} {\bibinfo  {journal} {Phys. Rev. Lett.}\ }\textbf {\bibinfo
  {volume} {81}},\ \bibinfo {pages} {5672} (\bibinfo {year}
  {1998})}\BibitemShut {NoStop}%
\bibitem [{\citenamefont {Datta}\ \emph {et~al.}(2008)\citenamefont {Datta},
  \citenamefont {Shaji},\ and\ \citenamefont {Caves}}]{Datta2008}%
  \BibitemOpen
  \bibfield  {author} {\bibinfo {author} {\bibfnamefont {A.}~\bibnamefont
  {Datta}}, \bibinfo {author} {\bibfnamefont {A.}~\bibnamefont {Shaji}}, \ and\
  \bibinfo {author} {\bibfnamefont {C.~M.}\ \bibnamefont {Caves}},\ }\href
  {\doibase 10.1103/PhysRevLett.100.050502} {\bibfield  {journal} {\bibinfo
  {journal} {Phys. Rev. Lett.}\ }\textbf {\bibinfo {volume} {100}},\ \bibinfo
  {pages} {050502} (\bibinfo {year} {2008})}\BibitemShut {NoStop}%
\bibitem [{\citenamefont {Bennett}\ \emph {et~al.}(1996)\citenamefont
  {Bennett}, \citenamefont {DiVincenzo}, \citenamefont {Smolin},\ and\
  \citenamefont {Wootters}}]{Bennett1996}%
  \BibitemOpen
  \bibfield  {author} {\bibinfo {author} {\bibfnamefont {C.~H.}\ \bibnamefont
  {Bennett}}, \bibinfo {author} {\bibfnamefont {D.~P.}\ \bibnamefont
  {DiVincenzo}}, \bibinfo {author} {\bibfnamefont {J.~A.}\ \bibnamefont
  {Smolin}}, \ and\ \bibinfo {author} {\bibfnamefont {W.~K.}\ \bibnamefont
  {Wootters}},\ }\href {\doibase 10.1103/PhysRevA.54.3824} {\bibfield
  {journal} {\bibinfo  {journal} {Phys. Rev. A}\ }\textbf {\bibinfo {volume}
  {54}},\ \bibinfo {pages} {3824} (\bibinfo {year} {1996})}\BibitemShut
  {NoStop}%
\bibitem [{Note1()}]{Note1}%
  \BibitemOpen
  \bibinfo {note} {We note that for general initial states $\rho ^{{\protect
  \cal SAR}}$ it is possible to create entanglement between ${\protect \cal S}$
  and ${\protect \cal R}$ via LOCC between ${\protect \cal A}$ and ${\protect
  \cal R}$. This can be demonstrated using the total initial state $\rho
  =\protect \frac {1}{2}\mathinner {|{0}\delimiter "526930B }\mathinner
  {\delimiter "426830A {0}|}^{{\protect \cal A}}\otimes \sigma ^{{\protect \cal
  SR}}+\protect \frac {1}{2}\mathinner {|{1}\delimiter "526930B }\mathinner
  {\delimiter "426830A {1}|}^{{\protect \cal A}}\otimes \tau ^{{\protect \cal
  SR}}$, which can have arbitrary little entanglement between ${\protect \cal
  S}$ and ${\protect \cal R}$, even for highly entangled states $\sigma $ and
  $\tau $. By LOCC between ${\protect \cal A}$ and ${\protect \cal R}$ it is
  possible to create the final state $\rho '=\protect \frac {1}{2}\mathinner
  {|{0}\delimiter "526930B }\mathinner {\delimiter "426830A {0}|}^{{\protect
  \cal R}'}\otimes \sigma ^{{\protect \cal SR}}+\protect \frac {1}{2}\mathinner
  {|{1}\delimiter "526930B }\mathinner {\delimiter "426830A {1}|}^{{\protect
  \cal R}'}\otimes \tau ^{{\protect \cal SR}}$, where the recipient is now in
  possession of ${\protect \cal R}$ and ${\protect \cal R}'$. Here, the
  entanglement between the system and the recipient can be large, even if they
  had little or even zero initial entanglement.}\BibitemShut {Stop}%
\bibitem [{\citenamefont {Horodecki}\ \emph {et~al.}(2003)\citenamefont
  {Horodecki}, \citenamefont {Shor},\ and\ \citenamefont
  {Ruskai}}]{Horodecki2003}%
  \BibitemOpen
  \bibfield  {author} {\bibinfo {author} {\bibfnamefont {M.}~\bibnamefont
  {Horodecki}}, \bibinfo {author} {\bibfnamefont {P.~W.}\ \bibnamefont {Shor}},
  \ and\ \bibinfo {author} {\bibfnamefont {M.~B.}\ \bibnamefont {Ruskai}},\
  }\href {\doibase 10.1142/S0129055X03001709} {\bibfield  {journal} {\bibinfo
  {journal} {Rev. Math. Phys.}\ }\textbf {\bibinfo {volume} {15}},\ \bibinfo
  {pages} {629} (\bibinfo {year} {2003})}\BibitemShut {NoStop}%
\bibitem [{\citenamefont {Vedral}(2002)}]{Vedral2002}%
  \BibitemOpen
  \bibfield  {author} {\bibinfo {author} {\bibfnamefont {V.}~\bibnamefont
  {Vedral}},\ }\href {\doibase 10.1103/RevModPhys.74.197} {\bibfield  {journal}
  {\bibinfo  {journal} {Rev. Mod. Phys.}\ }\textbf {\bibinfo {volume} {74}},\
  \bibinfo {pages} {197} (\bibinfo {year} {2002})}\BibitemShut {NoStop}%
\bibitem [{\citenamefont {Wheeler}\ and\ \citenamefont
  {Zurek}(1983)}]{Wheeler1983}%
  \BibitemOpen
  \bibinfo {editor} {\bibfnamefont {J.~A.}\ \bibnamefont {Wheeler}}\ and\
  \bibinfo {editor} {\bibfnamefont {W.~H.}\ \bibnamefont {Zurek}},\ eds.,\
  \href@noop {} {\emph {\bibinfo {title} {Quantum Theory and Measurement}}}\
  (\bibinfo  {publisher} {Princeton University Press},\ \bibinfo {address}
  {Princeton},\ \bibinfo {year} {1983})\BibitemShut {NoStop}%
\bibitem [{\citenamefont {Bennett}\ \emph {et~al.}(1993)\citenamefont
  {Bennett}, \citenamefont {Brassard}, \citenamefont {Cr\'epeau}, \citenamefont
  {Jozsa}, \citenamefont {Peres},\ and\ \citenamefont
  {Wootters}}]{Bennett1993}%
  \BibitemOpen
  \bibfield  {author} {\bibinfo {author} {\bibfnamefont {C.~H.}\ \bibnamefont
  {Bennett}}, \bibinfo {author} {\bibfnamefont {G.}~\bibnamefont {Brassard}},
  \bibinfo {author} {\bibfnamefont {C.}~\bibnamefont {Cr\'epeau}}, \bibinfo
  {author} {\bibfnamefont {R.}~\bibnamefont {Jozsa}}, \bibinfo {author}
  {\bibfnamefont {A.}~\bibnamefont {Peres}}, \ and\ \bibinfo {author}
  {\bibfnamefont {W.~K.}\ \bibnamefont {Wootters}},\ }\href {\doibase
  10.1103/PhysRevLett.70.1895} {\bibfield  {journal} {\bibinfo  {journal}
  {Phys. Rev. Lett.}\ }\textbf {\bibinfo {volume} {70}},\ \bibinfo {pages}
  {1895} (\bibinfo {year} {1993})}\BibitemShut {NoStop}%
\bibitem [{\citenamefont {Koashi}\ and\ \citenamefont
  {Winter}(2004)}]{Koashi2004}%
  \BibitemOpen
  \bibfield  {author} {\bibinfo {author} {\bibfnamefont {M.}~\bibnamefont
  {Koashi}}\ and\ \bibinfo {author} {\bibfnamefont {A.}~\bibnamefont
  {Winter}},\ }\href {\doibase 10.1103/PhysRevA.69.022309} {\bibfield
  {journal} {\bibinfo  {journal} {Phys. Rev. A}\ }\textbf {\bibinfo {volume}
  {69}},\ \bibinfo {pages} {022309} (\bibinfo {year} {2004})}\BibitemShut
  {NoStop}%
\bibitem [{\citenamefont {Bru\ss{}}\ \emph {et~al.}(1998)\citenamefont
  {Bru\ss{}}, \citenamefont {DiVincenzo}, \citenamefont {Ekert}, \citenamefont
  {Fuchs}, \citenamefont {Macchiavello},\ and\ \citenamefont
  {Smolin}}]{Bruss1998}%
  \BibitemOpen
  \bibfield  {author} {\bibinfo {author} {\bibfnamefont {D.}~\bibnamefont
  {Bru\ss{}}}, \bibinfo {author} {\bibfnamefont {D.~P.}\ \bibnamefont
  {DiVincenzo}}, \bibinfo {author} {\bibfnamefont {A.}~\bibnamefont {Ekert}},
  \bibinfo {author} {\bibfnamefont {C.~A.}\ \bibnamefont {Fuchs}}, \bibinfo
  {author} {\bibfnamefont {C.}~\bibnamefont {Macchiavello}}, \ and\ \bibinfo
  {author} {\bibfnamefont {J.~A.}\ \bibnamefont {Smolin}},\ }\href {\doibase
  10.1103/PhysRevA.57.2368} {\bibfield  {journal} {\bibinfo  {journal} {Phys.
  Rev. A}\ }\textbf {\bibinfo {volume} {57}},\ \bibinfo {pages} {2368}
  (\bibinfo {year} {1998})}\BibitemShut {NoStop}%
\bibitem [{\citenamefont {Wootters}\ and\ \citenamefont
  {Zurek}(1982)}]{Wootters1982}%
  \BibitemOpen
  \bibfield  {author} {\bibinfo {author} {\bibfnamefont {W.~K.}\ \bibnamefont
  {Wootters}}\ and\ \bibinfo {author} {\bibfnamefont {W.~H.}\ \bibnamefont
  {Zurek}},\ }\href {\doibase 10.1038/299802a0} {\bibfield  {journal} {\bibinfo
   {journal} {Nature (London)}\ }\textbf {\bibinfo {volume} {299}},\ \bibinfo
  {pages} {802} (\bibinfo {year} {1982})}\BibitemShut {NoStop}%
\bibitem [{\citenamefont {Dieks}(1982)}]{Dieks1982}%
  \BibitemOpen
  \bibfield  {author} {\bibinfo {author} {\bibfnamefont {D.}~\bibnamefont
  {Dieks}},\ }\href {\doibase 10.1016/0375-9601(82)90084-6} {\bibfield
  {journal} {\bibinfo  {journal} {Phys. Lett. A}\ }\textbf {\bibinfo {volume}
  {92}},\ \bibinfo {pages} {271 } (\bibinfo {year} {1982})}\BibitemShut
  {NoStop}%
\bibitem [{\citenamefont {Zurek}(2009)}]{Zurek2009}%
  \BibitemOpen
  \bibfield  {author} {\bibinfo {author} {\bibfnamefont {W.~H.}\ \bibnamefont
  {Zurek}},\ }\href {\doibase 10.1038/nphys1202} {\bibfield  {journal}
  {\bibinfo  {journal} {Nat. Phys.}\ }\textbf {\bibinfo {volume} {5}},\
  \bibinfo {pages} {181} (\bibinfo {year} {2009})}\BibitemShut {NoStop}%
\bibitem [{\citenamefont {Riedel}\ and\ \citenamefont
  {Zurek}(2010)}]{Riedel2010}%
  \BibitemOpen
  \bibfield  {author} {\bibinfo {author} {\bibfnamefont {C.~J.}\ \bibnamefont
  {Riedel}}\ and\ \bibinfo {author} {\bibfnamefont {W.~H.}\ \bibnamefont
  {Zurek}},\ }\href {\doibase 10.1103/PhysRevLett.105.020404} {\bibfield
  {journal} {\bibinfo  {journal} {Phys. Rev. Lett.}\ }\textbf {\bibinfo
  {volume} {105}},\ \bibinfo {pages} {020404} (\bibinfo {year}
  {2010})}\BibitemShut {NoStop}%
\bibitem [{\citenamefont {Zwolak}\ and\ \citenamefont
  {Zurek}(2013)}]{Zwolak2013}%
  \BibitemOpen
  \bibfield  {author} {\bibinfo {author} {\bibfnamefont {M.}~\bibnamefont
  {Zwolak}}\ and\ \bibinfo {author} {\bibfnamefont {W.~H.}\ \bibnamefont
  {Zurek}},\ }\href {\doibase 10.1038/srep01729} {\bibfield  {journal}
  {\bibinfo  {journal} {Sci. Rep.}\ }\textbf {\bibinfo {volume} {3}},\ \bibinfo
  {pages} {1729} (\bibinfo {year} {2013})}\BibitemShut {NoStop}%
\bibitem [{\citenamefont {Horodecki}\ \emph {et~al.}(2005)\citenamefont
  {Horodecki}, \citenamefont {Oppenheim},\ and\ \citenamefont
  {Winter}}]{Horodecki2005}%
  \BibitemOpen
  \bibfield  {author} {\bibinfo {author} {\bibfnamefont {M.}~\bibnamefont
  {Horodecki}}, \bibinfo {author} {\bibfnamefont {J.}~\bibnamefont
  {Oppenheim}}, \ and\ \bibinfo {author} {\bibfnamefont {A.}~\bibnamefont
  {Winter}},\ }\href {\doibase 10.1038/nature03909} {\bibfield  {journal}
  {\bibinfo  {journal} {Nature (London)}\ }\textbf {\bibinfo {volume} {436}},\
  \bibinfo {pages} {673} (\bibinfo {year} {2005})}\BibitemShut {NoStop}%
\bibitem [{\citenamefont {Piani}\ \emph {et~al.}(2008)\citenamefont {Piani},
  \citenamefont {Horodecki},\ and\ \citenamefont {Horodecki}}]{Piani2008}%
  \BibitemOpen
  \bibfield  {author} {\bibinfo {author} {\bibfnamefont {M.}~\bibnamefont
  {Piani}}, \bibinfo {author} {\bibfnamefont {P.}~\bibnamefont {Horodecki}}, \
  and\ \bibinfo {author} {\bibfnamefont {R.}~\bibnamefont {Horodecki}},\ }\href
  {\doibase 10.1103/PhysRevLett.100.090502} {\bibfield  {journal} {\bibinfo
  {journal} {Phys. Rev. Lett.}\ }\textbf {\bibinfo {volume} {100}},\ \bibinfo
  {pages} {090502} (\bibinfo {year} {2008})}\BibitemShut {NoStop}%
\bibitem [{\citenamefont {Luo}\ and\ \citenamefont {Sun}(2010)}]{Luo2010}%
  \BibitemOpen
  \bibfield  {author} {\bibinfo {author} {\bibfnamefont {S.}~\bibnamefont
  {Luo}}\ and\ \bibinfo {author} {\bibfnamefont {W.}~\bibnamefont {Sun}},\
  }\href {\doibase 10.1103/PhysRevA.82.012338} {\bibfield  {journal} {\bibinfo
  {journal} {Phys. Rev. A}\ }\textbf {\bibinfo {volume} {82}},\ \bibinfo
  {pages} {012338} (\bibinfo {year} {2010})}\BibitemShut {NoStop}%
\bibitem [{\citenamefont {Bohr}(1963)}]{Bohr1963}%
  \BibitemOpen
  \bibfield  {author} {\bibinfo {author} {\bibfnamefont {N.}~\bibnamefont
  {Bohr}},\ }\href@noop {} {\emph {\bibinfo {title} {Essays 1958 - 1962 on
  atomic physics and human knowledge}}}\ (\bibinfo  {publisher} {J. Wiley},\
  \bibinfo {address} {New York},\ \bibinfo {year} {1963})\BibitemShut {NoStop}%
\bibitem [{\citenamefont {Bae}\ and\ \citenamefont {Ac\'in}(2006)}]{Bae2006}%
  \BibitemOpen
  \bibfield  {author} {\bibinfo {author} {\bibfnamefont {J.}~\bibnamefont
  {Bae}}\ and\ \bibinfo {author} {\bibfnamefont {A.}~\bibnamefont {Ac\'in}},\
  }\href {\doibase 10.1103/PhysRevLett.97.030402} {\bibfield  {journal}
  {\bibinfo  {journal} {Phys. Rev. Lett.}\ }\textbf {\bibinfo {volume} {97}},\
  \bibinfo {pages} {030402} (\bibinfo {year} {2006})}\BibitemShut {NoStop}%
\bibitem [{\citenamefont {Chiribella}\ and\ \citenamefont
  {D'Ariano}(2006)}]{Chiribella2006}%
  \BibitemOpen
  \bibfield  {author} {\bibinfo {author} {\bibfnamefont {G.}~\bibnamefont
  {Chiribella}}\ and\ \bibinfo {author} {\bibfnamefont {G.~M.}\ \bibnamefont
  {D'Ariano}},\ }\href {\doibase 10.1103/PhysRevLett.97.250503} {\bibfield
  {journal} {\bibinfo  {journal} {Phys. Rev. Lett.}\ }\textbf {\bibinfo
  {volume} {97}},\ \bibinfo {pages} {250503} (\bibinfo {year}
  {2006})}\BibitemShut {NoStop}%
\bibitem [{\citenamefont {Brand\~ao}\ \emph {et~al.}()\citenamefont
  {Brand\~ao}, \citenamefont {Piani},\ and\ \citenamefont
  {Horodecki}}]{Brandao2013}%
  \BibitemOpen
  \bibfield  {author} {\bibinfo {author} {\bibfnamefont {F.~G. S.~L.}\
  \bibnamefont {Brand\~ao}}, \bibinfo {author} {\bibfnamefont {M.}~\bibnamefont
  {Piani}}, \ and\ \bibinfo {author} {\bibfnamefont {P.}~\bibnamefont
  {Horodecki}},\ }\href@noop {} {}\bibinfo {note} {{in
  preparation}}\BibitemShut {NoStop}%
\end{thebibliography}%

\end{document}